\documentclass[letterpaper, 10 pt, conference]{ieeeconf}  

\IEEEoverridecommandlockouts                              

\usepackage{graphics} 
\usepackage{graphicx}
\usepackage{epsfig} 
\usepackage{mathptmx} 
\usepackage{times} 
\usepackage{amsmath} 
\usepackage{amssymb}  
\usepackage{mathtools}
\usepackage{euscript} 
\usepackage{hyperref}
\usepackage{xurl}

\usepackage{cite}

\usepackage{color}
\usepackage{balance}

\newtheorem{definition}{Definition}

\newtheorem{remark}{\bf Remark}
\newtheorem{assumption}{Assumption}
\newtheorem{theorem}{Theorem}

\newtheorem{proposition}{Proposition}

\DeclareGraphicsExtensions{.png,.jpg,.eps}

\title{\LARGE \bf
Learning Neural Koopman Operators with Dissipativity Guarantees
}

\author{Yuezhu Xu, S. Sivaranjani, Vijay Gupta 
\thanks{Y. Xu and S. Sivaranjani are with the Edwardson School of Industrial Engineering, and V. Gupta is with the Elmore Family School of Electrical and Computer Engineering, at Purdue University, West Lafayette, IN, USA.
       \{{\tt\small xu1732,sseetha,gupta869\}}{\tt\small@purdue.edu}. { This work was partially supported by the Air Force Office of Scientific
Research grant, FA9550-23-1-0492.
}
}
}

\begin{document}

\maketitle
\thispagestyle{empty}
\pagestyle{empty}

\begin{abstract}
We address the problem of learning a neural Koopman operator model that provides dissipativity guarantees for an unknown nonlinear dynamical system that is known to be dissipative. We propose a two-stage approach. First, we learn an unconstrained neural Koopman model that closely approximates the system dynamics. Then, we minimally perturb the parameters to enforce strict dissipativity. Crucially, we establish theoretical guarantees that extend the dissipativity properties of the learned model back to the original nonlinear system. We realize this by deriving an exact relationship between the dissipativity of the learned model and the true system through careful characterization of the identification errors from the noisy data, Koopman operator truncation, and generalization to unseen data. We demonstrate our approach through simulation on a Duffing oscillator model. 
\end{abstract}

\section{Introduction}\label{sec: intro}
Nonlinear system identification is a topic of central interest in control, and has seen renewed interest with recent developments in data-driven approaches such as Koopman operators \cite{proctor2016dynamic,williams2015data,brunton2016discovering}, and deep learning based models such as neural ODEs \cite{ljung2020deep,yu2024learning}. As with any system identification problem,  \textit{identification for control} -- where the goal is to obtain closed-loop guarantees when the learned models are used for control design \cite{gevers2005identification} -- is an important consideration. One approach to ensure that the guarantees from the learned open-loop model can be ported to guarantees on the closed-loop system is to incorporate a priori information on control-relevant properties such as stability or passivity into the learned model, and exploit them during control design to obtain closed-loop guarantees. Works on the problem of preserving control-relevant properties in system identification have largely focused on linear systems, where stability \cite{pintelon2012system}, positive-realness \cite{goethals2003identification,hoagg2004first}, passivity \cite{grivet2015passive}, and dissipativity \cite{sivaranjani2022data} have been imposed on the learned models. For preserving control-relevant properties in nonlinear identification, there has been some very recent work on learning Koopman operator based models that are provably stable \cite{fan2022learning,bevanda2022diffeomorphically,wang2024optimal}, negative imaginary \cite{mabrok2023koopman}, or dissipative  \cite{hara2020learning}, and neural ODE based models with Lyapunov stability \cite{kolter2019learning} and dissipativity \cite{xu2023learning,drgovna2022dissipative} guarantees. 

However, ensuring that a learned model  satisfies a particular property, such as stability, is different from establishing that the same properties of the learned model reflect that of the true and unknown system. Arguably, the latter question is more relevant to provide closed-loop guarantees in control designs that exploit these properties. Consider an unknown system that is known to be dissipative. Even if existing techniques can ensure that a model of this system identified from data is guaranteed to be dissipative with certain indices, controllers synthesized from the learned model that rely on these indices do not necessarily yield the closed-loop guarantees when applied to the original system.
In other words, in the absence of a certified relationship between the true system and the learned model, a simple enforcement that the learned model be dissipative may not provide guarantees for the true system when the learned model is used to design a controller. Closing this gap is crucial in providing closed-loop guarantees using learned models. 

In this paper, we consider the problem of learning a Koopman operator model for an unknown nonlinear dynamical system that is a priori known to be dissipative. In addition to learning a model that closely approximates the system dynamics, we would like to  guarantee that the learned model preserves the dissipativity property of the nonlinear system. Further, we would like to derive a relationship between the dissipativity of the learned model and that of the true system. The latter is crucial in preserving closed-loop guarantees when the learned model is utilized for designing controllers that are deployed on the original unknown system. We focus on dissipativity since it is an important control property encompassing  special cases such as $\mathcal{L}_2$ stability, passivity, and sector-boundedness \cite{brogliato2007dissipative} and is commonly leveraged in distributed and compositional control designs \cite{Tippett2013,antsaklis2013control,Agarwal2021}.

There are several approaches to identify Koopman operator models in both open-loop and closed-loop settings  as well as data-driven methods to learn finite-dimensional linear representations of the Koopman operator (we refer the reader to works such as~\cite{williams2015data,mauroy2016linear,mauroy2020koopman,bevanda2021koopman,proctor2018generalizing,korda2018linear,proctor2016dynamic} and the references therein for an overview). The Koopman operator lifts the system states to observables in a high-dimensional space where we can learn a linear representation of the dynamics.  Traditional methods typically use heuristically pre-determined
dictionaries of observables, which makes them sensitive to the choice of lifting functions, and often necessitates manual tuning. 
Deep learning methods have been proposed to mitigate this, learning embeddings that yield globally linear models directly from data \cite{lusch2018deep,hao2024deep,meng2024koopman}. In parallel, recent works show that for systems with inputs, the lifted representation may take bilinear or LPV forms \cite{iacob2024koopman}, and finite-data error bounds have been established for such models \cite{nuske2023finite}. In this paper, we focus on linear lifted models and demonstrate how dissipativity can be enforced within this setting by learning \textit{neural Koopman models}, that is, finite-dimensional, data-driven surrogates of the Koopman operator where deep neural networks are used to learn the observables.

One common approach to preserve control-relevant properties in system identification  is to impose explicit constraints during the training process \cite{hara2020learning}. However, extending such approaches to learn neural Koopman operators requires constrained neural network training, which typically requires computationally expensive projections \cite{kolter2019learning,okamoto2024learning}. Constrained identification is also not desirable in general, even in linear models, as it introduces a trade-off between optimal data noise filtering and model constraints, resulting in uncontrollable model bias \cite{pintelon2012system}. We address this challenge through a two-stage approach. First, we learn an unconstrained neural Koopman operator that closely approximates the nonlinear system dynamics. Next, we design linear matrix inequalities (LMI)-based parameter perturbations to enforce strict dissipativity, while minimizing deviation from the unconstrained model. Crucially, we design these perturbations to ensure that strict dissipativity of the model translates to dissipativity guarantees on the original nonlinear system. Specifically, we derive an exact relationship between the dissipativity of the learned model and the true system by characterizing and bounding errors across the entire identification process, including errors due to noisy data, Koopman operator identification and truncation, and generalization to unseen data.

We note that existing works \cite{mabrok2023koopman,hara2020learning}  do not provide any guarantees on the dissipativity of the original unknown system based on the learned model, which is a \textit{key contribution} of this work. We also note that perturbation approaches have been employed to enforce dissipativity-type properties \cite{grivet2015passive,sivaranjani2022data,xu2023learning}; however, in these works, the dissipativity guarantees on the learned model do not extend back to the true system beyond the linear near-equilibrium region.
In other words, we cannot obtain guarantees on the true system  from the learned model through such methods. In contrast, our approach extends such guarantees to the original nonlinear system that may be operating  far from equilibrium.  
This paper is organized as follows. We provide preliminaries on the Koopman operator  and dissipativity in Section \ref{sec: problem}, detail our approach to learn dissipative neural Koopman operator models in Section \ref{sec: learning} and validate its performance through simulation on a Duffing oscillator system in Section \ref{sec: case}. All the proofs are presented in the Appendix. 

\textbf{Notation:}
We denote the space of all real, non-negative real, and positive real numbers, $n$-dimensional real vectors, and non-negative integers by $\mathbb{R}$, $\mathbb{R}_{+}$, $\mathbb{R}_{++}$, $\mathbb{R}^{n}$, and $\mathbb{Z}_+$ respectively. $A'$ represents the transpose of matrix $A$. A symmetric matrix \( P \in \mathbb{R}^{n \times n} \) is positive definite if \( P > 0 \) (or positive semidefinite if \( P \ge 0 \)) and negative definite if \( P < 0 \) (or negative semidefinite if \( P \le 0 \)). The identity matrix is denoted by ${I}$ with dimensions clear from context. $\|\cdot\|$ and $\|\cdot\|_F$ represent the 2-norm and Frobenius norm respectively.

\section{Preliminaries} \label{sec: problem}
Consider a nonlinear discrete time-invariant system 
\begin{equation}\label{eqn: nonlinear dynamics}
\begin{aligned} 
\Sigma_{nl}: \quad     x(k+1) = f(x(k), u(k)), \quad    y(k) = g(x(k),u(k)),
\end{aligned}
\end{equation}
where  $x(k)\in  \mathcal{X} 
\subset \mathbb{R}^n$, $u(k)\in\mathcal{U}\subset\mathbb{R}^m$, and $y(k)\in\mathcal{Y}\subset\mathbb{R}^p$ are the state, input, and output vectors at time $k\in \mathbb{Z}_{+}$ respectively, and $f: \mathcal{X}\times \mathcal{U}\rightarrow \mathcal{X}$ and $g:\mathcal{X}\times \mathcal{U}\rightarrow \mathcal{Y}$,  with $\mathcal{X}$ and $\mathcal{U}$ bounded. Here $\mathcal{X}$ is the state domain under consideration. Note that we do not require additional assumptions such as forward invariance of $\mathcal{X}$.

\begin{assumption} \label{assp: Lip on f and g}
\begingroup
\setlength{\abovedisplayskip}{5pt}
\setlength{\belowdisplayskip}{2pt}
    The functions $f$ and $g$  are Lipschitz continuous with respect to $x$, with Lipschitz constants  $L_f^x$ and $L_g^x$ respectively. Thus, for any $x_1,x_2\in\mathcal{X}$ and $u\in\mathcal{U}$, we have
    \begin{equation} \label{eqn: Lip on f and g}
        \begin{aligned}
            &\|f(x_1,u)-f(x_2,u)\|\leq L_f^x\|x_1-x_2\|\\
            & \|g(x_1,u)-g(x_2,u)\|\leq L_g^x\|x_1-x_2\|.
        \end{aligned}
    \end{equation}
\endgroup
\end{assumption}

\begin{definition}[Dissipativity / Strict Dissipativity]\label{def: qsr_dissipativity} Suppose that for any $u(k)$ and $y(k)$, and matrices $Q,S,R$ of appropriate dimensions where $Q\leq0$, there exists a storage function $V(x(k))$ such that for any $k\in\mathbb{R}_+$
\begin{equation} \label{eqn: strict QSR}
    V(x(k+1)) - V(x(k)) \leq w({\color{blue}y}(k), u(k)) - \alpha({\color{blue}y}(k), u(k)),
\end{equation}
where 
\begin{align*}
    w(y(k), u(k))&:=y'(k)Qy(k)+2y'(k)Su(k)+u'(k)Ru(k)\\
    \alpha(x(k), u(k)) &:= \rho x'(k)x(k)+\nu u'(k)u'(k).
\end{align*}
The system $\Sigma_{nl}$ is \textit{strictly $(Q,S,R)$-dissipative} with supply rate $w(y(k),u(k))$ if~(\ref{eqn: strict QSR}) holds with $\rho>0,\nu>0$ and \textit{$(Q,S,R)$-dissipative} (or simply dissipative) if~(\ref{eqn: strict QSR}) holds with $\rho=0,\nu=0$.
\end{definition}
Appropriate choices of the matrices $(Q,S,R)$ can capture a broad range of control-relevant properties with notable special cases including $\mathcal{L}_2$ stability ($Q = -\frac{1}{\gamma} I, S = 0, R=\gamma I$, $\gamma > 0$), passivity ($Q =0, S=\frac{1}{2}I, R=0$), strict Passivity ($Q = -\epsilon I, S=\frac{1}{2}I, R=-\delta I$, $\epsilon, \delta>0$), and sector-boundedness ($Q=-I$, $S = (a+b)I$, $R = -ab I$.)
  \begin{assumption}\label{assp: disspativity_nl}
    The system $\Sigma_{nl}$ is $(Q,S,R)$-dissipative.
\end{assumption}

It is desirable to learn a linear representation of the dynamics of $\Sigma_{nl}$ beyond the typical near-equilibrium operating region where linearized models obtained using Taylor expansion are applicable. One popular approach is to use the Koopman operator, in which the states $x\in\mathbb{R}^n$ are lifted to an infinite-dimensional {\em observable} $z_{inf}$. In the lifted space, the dynamics of $z_{inf}$ evolves as a linear system.
In practice, the infinite-dimensional $z_{inf}$ is truncated to a finite but high-dimensional $z \in \mathbb{R}^N,$ with $N\gg n$, where $z=\phi(x)$, and  $\phi: \mathbb{R}^n\rightarrow\mathbb{R}^N$  is a finite-dimensional \textit{lifting function}.
Denote the resulting high-dimensional linear system by
\begin{equation} \label{eqn: linear dynamics}
    \Sigma_l: \;    z(k+1) = Az(k)+Bu(k),  \quad
         \tilde{y}(k) = Cz(k)+Du(k), 
\end{equation}
where $\tilde{y}$ is the approximated output in the truncated space. 
\begin{assumption} \label{assp: input}
    We assume that the input $u(k)$ is known.
\end{assumption}
Assumption \ref{assp: input} is justified as it is commonplace to design or select suitable  inputs to facilitate accurate identification. 

Given that $\Sigma_{nl}$ is known a priori to be dissipative (Assumption \ref{assp: disspativity_nl}), our goal is to learn a  Koopman operator model $\Sigma_l$ that closely approximates the state trajectories and input-output behavior of $\Sigma_{nl}$ while also being dissipative. We note that dissipativity is not inherently preserved in any learned model, even if the model closely approximates the system  dynamics \cite{pintelon2012system}. Further, dynamical system data are typically noisy due to measurement uncertainties or disturbances, and there are identification errors in learning the  Koopman operator, and model reduction errors due to the finite-dimensional truncation of the Koopman operator. 
These compounding errors imply that even if we explicitly learn a dissipative model, we cannot extend dissipativity guarantees back to the original nonlinear system. Thus, when the learned model is used in control design, there is no guarantee that the closed-loop dissipativity guarantees on the learned model will also hold for the nonlinear system. Therefore, we also aim to derive conditions under which  dissipativity guarantees on $\Sigma_l$  translate to dissipativity guarantees on $\Sigma_{nl}$.

\section{Learning Dissipative \\Neural Koopman Operators}\label{sec: learning}
\subsection{Learning Neural Koopman Operators} \label{sec: learning neural Koopman}
Data-driven methods approaches to learn Koopman operator representations of a dynamical system \cite{williams2015data,korda2018linear} typically use heuristically predetermined dictionaries of observables. However, the choice of lifting functions can significantly impact the accuracy of the learned model, and selecting an optimal dictionary is non-trivial.
To overcome this limitation, we utilize neural Koopman methods and leverage deep neural networks to automatically learn the observables.  Specifically, we model the lifting function and the projection from observables back to the state space, which we term as the \textit{encoder} $\phi$ and \textit{decoder} $\psi$ respectively, as
$z = \phi(x)$ {and}  $\tilde{x} = \psi(z),$
where $\tilde{x}$ is the reconstructed state, and $\phi,\psi$ are modeled by two separate $L$ layer feedforward neural networks represented as 
   $ \phi = f_L^\phi \circ f_{L-1}^\phi \circ \dots \circ f_1^\phi,$ and $ 
\psi = f_L^\psi \circ f_{L-1}^\psi \circ \dots \circ f_1^\psi,$
where
    $f_i^\phi(u) = \sigma(W_i^\phi u + b_i^\phi),$ and $f_i^\psi(v) = \sigma(W_i^\psi v + b_i^\psi).$
Here, $W_i^\phi, W_i^\psi$ and $b_i^\phi, b_i^\psi$ are the weights and biases for $\phi$ and $\psi$ respectively, and $\sigma(\cdot)$ is an element-wise nonlinear activation function. We collect training data corresponding to $d$ trajectories each of length $M^{tr}$ from system $\Sigma_{nl}$, where data from the $i$-{th} trajectory are represented as
\begin{equation} \label{data train}
    \mathcal{D}^{tr}_i = \{(\hat{x}^{tr}_i(k), u^{tr}_i(k), \hat{x}^{tr}_i(k+1),\hat{y}^{tr}_i(k))\}_{k=1}^{M^{tr}}.
\end{equation}
We also denote the lifted states from $\hat{x}_i^{tr}(k)$ to  be $z_i(k)$.   In practice, training data collected from dynamical systems is noisy; we require that these noises be  small in the following sense to allow accurate identification.
\begin{assumption} \label{assp: noise bound}
   The distance between the collected trajectory data in \eqref{data train} and the true states and outputs $x(k)$ and $y(k)$ respectively, is bounded by a small constant $\delta_d > 0$, that is,
        $\|\hat{x}^{tr}_i(k)-x(k)\|\leq \delta_d,$ and $\|\hat{y}^{tr}_i(k)-y(k)\|\leq \delta_d, \forall k,i$.
\end{assumption} 

To learn the neural Koopman model, we define a three-part loss function consisting of the Koopman dynamic loss $\mathcal{L}_{z}$, prediction loss $\mathcal{L}_{pred}$, and  reconstruction loss $\mathcal{L}_{rec}$, where

\begin{equation*}
   {\small \begin{aligned}
        \mathcal{L}_{z}&\triangleq \frac{1}{M^{tr}d}\sum\limits_{i=1}^d \sum\limits_{k=1}^d\|z_i(k+1)-Az_i(k)-Bu^{tr}_i(k)\|\\
        & = \frac{1}{M^{tr}d}\sum\limits_{i=1}^d \sum\limits_{k=1}^d\|\phi(\hat{x}^{tr}_i(k+1))-A\phi(\hat{x}^{tr}_i(k))-Bu^{tr}_i(k)\|\\
        \mathcal{L}_{pred}& \triangleq \frac{1}{M^{tr}d}\sum\limits_{i=1}^d \sum\limits_{k=1}^d\|\hat{y}^{tr}_i(k)-Cz_i(k)-Du^{tr}_i(k)\|
    \end{aligned}}
\end{equation*}
\begin{equation*}
    \begin{aligned}
        \mathcal{L}_{rec}& \triangleq \frac{1}{M^{tr}d}\sum\limits_{i=1}^d \sum\limits_{k=1}^d\|\tilde{x}_i(k)-\hat{x}^{tr}_i(k)\|\\
        & =\frac{1}{M^{tr}d}\sum\limits_{i=1}^d \sum\limits_{k=1}^d\|\psi(\phi(\hat{x}^{tr}_i(k)))-\hat{x}^{tr}_i(k)\|.
    \end{aligned}
\end{equation*}

Then, the loss function can be expressed as 
\begin{equation} \label{eqn: Koopman loss}
    \mathcal{L} = \lambda_1\mathcal{L}_{z}+\lambda_2\mathcal{L}_{pred}+\lambda_3\mathcal{L}_{rec},
\end{equation}
where $\lambda_1, \lambda_2,\lambda_3\in \mathbb{R}_{++}$ are weighting parameters. Minimizing $\mathcal{L}$ on the training data \eqref{data train} allows us to effectively learn  observables and project them back  to accurately reconstruct the states while capturing the system dynamics.
\subsection{Dissipativity Guarantee on the Nonlinear System}
As discussed in Sec \ref{sec: problem}, dissipativity of $\Sigma_{nl}$ does not guarantee dissipativity of $\Sigma_l$ and vice-versa. 
We now derive a relationship between the dissipativity properties of $\Sigma_{nl}$ and the strict dissipativity of $\Sigma_l$ by bounding the errors arising from data collection, model identification, and generalization to unseen data. We have already bounded data collection noises through Assumption \ref{assp: noise bound}. We evaluate the discrepancy between the model-generated data and the collected data through a separate validation process. We collect $M^{ev}$ evaluation data tuples   denoted as 
\begin{equation} \label{data evaluation}
    \mathcal{D}^{ev} = \{(\hat{x}_j^{ev}, u_j^{ev}, \hat{y}_j^{ev}\}_{j=1}^{M^{ev}},
    \end{equation}
    where $\hat x^{ev}_j$ and $\hat y^{ev}_j$ are system state and output data corresponding to evaluation input $u^{ev}_j$ collected from system $\Sigma_{nl}.$
{\begin{assumption} \label{assp: data density}
    The collected state data in \eqref{data evaluation} is \textit{sufficiently dense} in the sense that for any $x_0 \in \mathcal{X}$, there exists a collected data point $\hat{x}^{ev}_j$, $j\in\{1,2,...M^{ev}\}$, satisfying
         $\|\hat{x}^{ev}_j-x_0\|\leq \delta_x$, where $\delta_x>0$ is a small constant.
\end{assumption}}
With this evaluation dataset, we define errors
\begin{equation} \label{eqn: delta_bd,delta_r} 
\begin{aligned}
     \delta_{c}\triangleq \max\limits_{j\in \{1,2,...,M^{ev}\}} \|\tilde{x}_j-\hat{x}^{ev}_j\|, \quad 
     \delta_{r}\triangleq \max\limits_{j\in \{1,2,...,M^{ev}\}} \|\tilde{y}_j-\hat{y}^{ev}_j\|,
\end{aligned}
\end{equation}
where $\tilde{x}_j=\psi(z_j)$, and $z_j$ and $\tilde{y}_j$ are the states and outputs obtained from the model for evaluation input $u^{ev}_j.$
Given sufficiently dense data collection (Assumption \ref{assp: data density}) in the validation process, and the encoder, decoder, and output functions being Lipschitz continuous, we can  upper bound the model error for unseen data through their corresponding Lipschitz constants. 
We have the following result relating the dissipativity of $\Sigma_l$ and $\Sigma_{nl}$.
\begin{theorem}\label{thm: dissipativity}
   If there exist $P=P'>0$, $P\in\mathbb{R}^l$ and $\rho,\nu>0$, such that 
      \begin{equation} \label{ineq: strict dissipative}
      {\small
        \begin{bmatrix}
        A' PA - P - C' Q C + \rho I & A'P B - C' QD - C' S \\[6pt]
        B' P A - S' C - D' Q C & B'PB -D'QD  - D' S - S' D -R + \nu I
    \end{bmatrix}\leq 0,}
   \end{equation}
then $\Sigma_{l}$ is strictly dissipative with indices $(Q,S,R,\rho, \nu)$.   Furthermore, for $\eta>0$, if there exist $\rho,\nu$ satisfying 
   \begin{subequations}
 {\small \label{ineq: rho, nu}
   \begin{align} 
       \rho\geq& \frac{(L^z_{\psi})^2}{(1-\delta_b/\eta-\|\psi(0)\|/\eta)^2}\times
       \label{ineq: rho}\\       &\Bigg[\frac{\delta_g(C)^2(\|Q\|+\tau+1)+\delta_g(C)\|Q\|\|C\|\|\phi(0)\|}{\eta^2}\\
       &+\frac{2\delta_g(C)\|Q\|\|C\|L^x_{\phi}}{\eta}\Bigg],\notag\\
       \nu\geq &\|Q\|^2\|D\|^2+\|S\|^2/\tau,
       \label{ineq: nu}
   \end{align}}
   \end{subequations}
where $\delta_g(C)\triangleq \delta_d+\delta_r+L^x_g\delta_x+L^x_{\phi}\delta_x\|C\|$ and $\delta_b\triangleq L^x_{\psi\circ\phi}\delta_x+\delta_{c}+\delta_d+\delta_x$, then the corresponding  $\Sigma_{nl}$ is guaranteed to be $(Q,S,R)$-dissipative   for $\|x\|\geq\eta>0$.
   Here $L^x_\phi$, $L^z_\psi$, $L^x_{\psi\circ\phi}$, and $L^x_g$ represent the Lipschitz constants of the functions $\phi$, $\psi$, $\psi\circ\phi$ and $g$ respectively, $\delta_d$ and $\delta_x$ are defined in Assumption \ref{assp: noise bound} and \ref{assp: data density} respectively, $\delta_{c}$ and $\delta_r$ are defined in \eqref{eqn: delta_bd,delta_r}, and $\tau>0$ is a user-defined hyper-parameter.
\end{theorem}
\begin{remark}
The choice of the parameter $\eta$ can be used to trade off the region where dissipativity is guaranteed against the scale of the perturbation. Specifically, a larger value of $\eta$ extends dissipativity guarantees farther from equilibrium but makes the condition on $\rho$ to enforce dissipativity more demanding. Note that dissipativity is particularly important when the system is operating far from equilibrium and we cannot obtain guarantees like stability through classical linear system analysis.
\end{remark}

While the proof of Theorem \ref{thm: dissipativity} is deferred to the Appendix, we outline the intuition here. With good initial training closely approximating the system dynamics, $\delta_r$ and $\delta_c$ are small. Additionally, given small noise levels (Assumption \ref{assp: noise bound}), dense data collection (that is, a small $\delta_x$), and reasonable Lipschitz constants, both $\delta_b$ and $\delta_g(C)$ remain small, where $\delta_g(C)$ consists of a small constant term and a small coefficient that scales with $\|C\|$. Then, the inequalities \eqref{ineq: rho, nu} are minimal additional requirements imposed beyond the standard dissipativity condition to guarantee that the dissipativity guarantee on $\Sigma_l$ extends back to the unknown system $\Sigma_{nl}$. The parameter $\tau$ is a user-defined hyperparameter to tune the trade-off between the error-induced constraints on $\rho$ and $\nu$. Finally, by choosing an appropriate $\eta$, we ensure dissipativity in the far-from-equilibrium region of the state space where  such guarantees are particularly difficult to obtain.

\subsection{Perturbation Approach and Practical  Considerations}\label{sec: alg}
While a common approach to preserving control-relevant properties like dissipativity in system identification is to  explicitly enforce constraints during training, this introduces two issues here, namely, difficulties in imposing hard constraints during neural network training, and uncontrollable model bias induced by model constraints \cite{pintelon2012system}.
We propose a two-stage approach to learn a neural Koopman model $\Sigma_l$ that approximates  $\Sigma_{nl}$ with dissipativity guarantees. First, we learn an unconstrained neural Koopman operator model using the objective function \eqref{eqn: Koopman loss} with data collection \eqref{data train} as in Section \ref{sec: learning neural Koopman}. 
From this process, we obtain encoder $\phi$, decoder $\psi$ and initial system matrices $(A_{ini}, B_{ini}, C_{ini}, D_{ini})$. We fix the parameters for encoder and decoder and 
sample densely to obtain evaluation data \eqref{data evaluation} on $\mathcal{X}$ such that $\delta_r$ is small, and record $\delta_{c}$ and $\delta_r$ as in \eqref{eqn: delta_bd,delta_r}. We also estimate the Lipschitz constant $L_g^x$ as defined in \eqref{eqn: Lip on f and g} by densely sampling on $\mathcal{X}$ and keeping $u$ constant. The Lipschitz constants of the neural networks  $\phi$, $\psi$, and $\psi\circ\phi$ are estimated using the  \textit{ECLipsE} algorithm \cite{xu2025eclipse}. Next, we will  minimally perturb the $(A_{ini},B_{ini},C_{ini},D_{ini})$ matrices of the baseline model  to impose constraint \eqref{ineq: strict dissipative}. Importantly, we will design  this perturbation such that strict dissipativity of $\Sigma_l$ guarantees the dissipativity of $\Sigma_{nl},$ as in Theorem \ref{thm: dissipativity}. Notice that \eqref{ineq: strict dissipative} is non-convex with respect to decision variables $A,B,C,D,P$ and the inequalities \eqref{ineq: rho, nu} are also quadratic with respect to $\|C\|$
and $\|D\|$. Such constraints make the problem numerically intractable. Therefore we transform \eqref{ineq: strict dissipative} and \eqref{ineq: rho, nu} into LMIs as follows. 
\begin{proposition} \label{thm: LMI}
Define  $X\triangleq PA$, $Y\triangleq PB$, $\delta_1\triangleq \delta_d+\delta_r+L^x_g\delta_x, \delta_2\triangleq L^x_\phi\delta_x$,  $c_\phi\triangleq(2\eta L^x_\phi+\|\phi(0)\|)$, and $N_C, N_D$ be slack variables of suitable dimensions. Then, for $Q<0$, \eqref{ineq: strict dissipative}-\eqref{ineq: rho, nu} are equivalent to the following LMIs: 
\begin{subequations} \label{LMI: Q<0}
    {\small
    \begin{gather}
        \begin{bmatrix}
               P-\rho I & C'S & C' & X'\\
               S'C & D'S+S'D+R-\nu I & D' & Y' \\
               C & D & -Q^{-1} & 0\\
               X & Y & 0 & P
        \end{bmatrix} \geq 0
        \label{LMI: Q<0a} \\
        \begin{bmatrix}
                \rho + \frac{(L^z_\psi)^2c_\phi\|Q\|\delta_1}{(\eta-\delta_b-\|\psi(0)\|)^2\delta_2}(\delta_1+\delta_2N_C) & \delta_1+\delta_2N_C  \\ 
                \delta_1+\delta_2N_C & \frac{(\eta-\delta_b-\|\psi(0)\|)^2\delta_2}{(L^z_\psi)^2((\|Q\|+\tau+1)\delta_2+\|Q\|c_\phi)}
        \end{bmatrix} \geq 0
        \label{LMI: Q<0b} \\
        \begin{bmatrix}
                \nu I-\|S\|^2/\tau & N_D \\ 
                N_D & \|Q\|^{-1}
        \end{bmatrix} \geq 0]
        \label{LMI: Q<0c} \\
        \|C\|\leq N_C, \quad \|D\|\leq N_D.
        \label{LMI: Q<0d}
    \end{gather}}
\end{subequations}
    For $Q=0$, \eqref{ineq: strict dissipative}-\eqref{ineq: rho, nu} are equivalent to 
        \begin{subequations}\label{LMI: Q=0}
        {\small
        \begin{gather}
            \begin{bmatrix}
               P-\rho I& C'S & X'\\
               S'C & D'S+S'D+R-\nu I & Y' \\
               X & Y  & P
            \end{bmatrix}\geq0
            \label{LMI: Q=0a}\\
            \begin{bmatrix}
                \rho  & \delta_1+\delta_2N_C  \\ 
                \delta_1+\delta_2N_C & \frac{(\eta-\delta_b-\|\psi(0)\|)^2}{(L^z_\psi)^2(\tau+1)} 
            \end{bmatrix}\geq0 
            \label{LMI: Q=0b}\\
            \nu\geq\|S\|^2/\tau, \quad \|C\|\leq N_C. \label{LMI: Q=0c}
        \end{gather}}
    \end{subequations}
\end{proposition}

We now solve the following optimization problem to perturb the initial system matrices to enforce dissipativity.
\begin{equation} \label{alg: perturbation}
\begin{aligned}
     \min_{X,Y,C,D,P,\rho,\nu,N_C,N_D} \quad & \lambda_A\|X - PA_{\text{ini}}\|_F^2 + \lambda_B \|Y - PB_{\text{ini}}\|_F^2 \\
     &+ \lambda_C \|C - C_{\text{ini}}\|_F^2 + \lambda_D\|D - D_{\text{ini}}\|_F^2 \\
    \text{s.t.} \quad &
    \begin{cases}
        \text{if } Q < 0, \quad \text{\eqref{LMI: Q<0} holds} & \\
        \text{if } Q = 0, \quad \text{\eqref{LMI: Q=0} holds}. & 
    \end{cases}
\end{aligned}
\end{equation}

\begin{remark}
It is worth noting that when the first-stage model already fits the system well and the true system is dissipative, the required perturbation is expected to be small, since the constraints in Theorem \ref{thm: LMI} are not demanding (the $\delta$s are very small in practice), thereby preserving both low validation error and the practical value of the learned model.
\end{remark}

With the optimal solution $(\bar{X},\bar{Y},\bar{C},\bar{D}, \bar{P},\bar{\rho},\bar{\nu},\bar{N}_C,\bar{N}_D)$, we calculate $A=\bar{P}^{-1}\bar{X}$, $B=\bar{P}^{-1}\bar{Y}$, $C=\bar{C}$, and $D=\bar{D}$. At this stage, we achieve our final dissipative neural Koopman model, with encoder $\phi$, decoder $\psi$, and lifted space dynamics $({A},{B},{C},{D})$ as in \eqref{eqn: linear dynamics}, which captures the nonlinear dynamics while extending  dissipativity guarantees from the lifted space back to the original nonlinear system.

\section{Case Study}\label{sec: case}
\begin{figure}[t] 
      \centering
      \includegraphics[scale=0.35,trim=0cm 1.2cm 0cm 0cm]{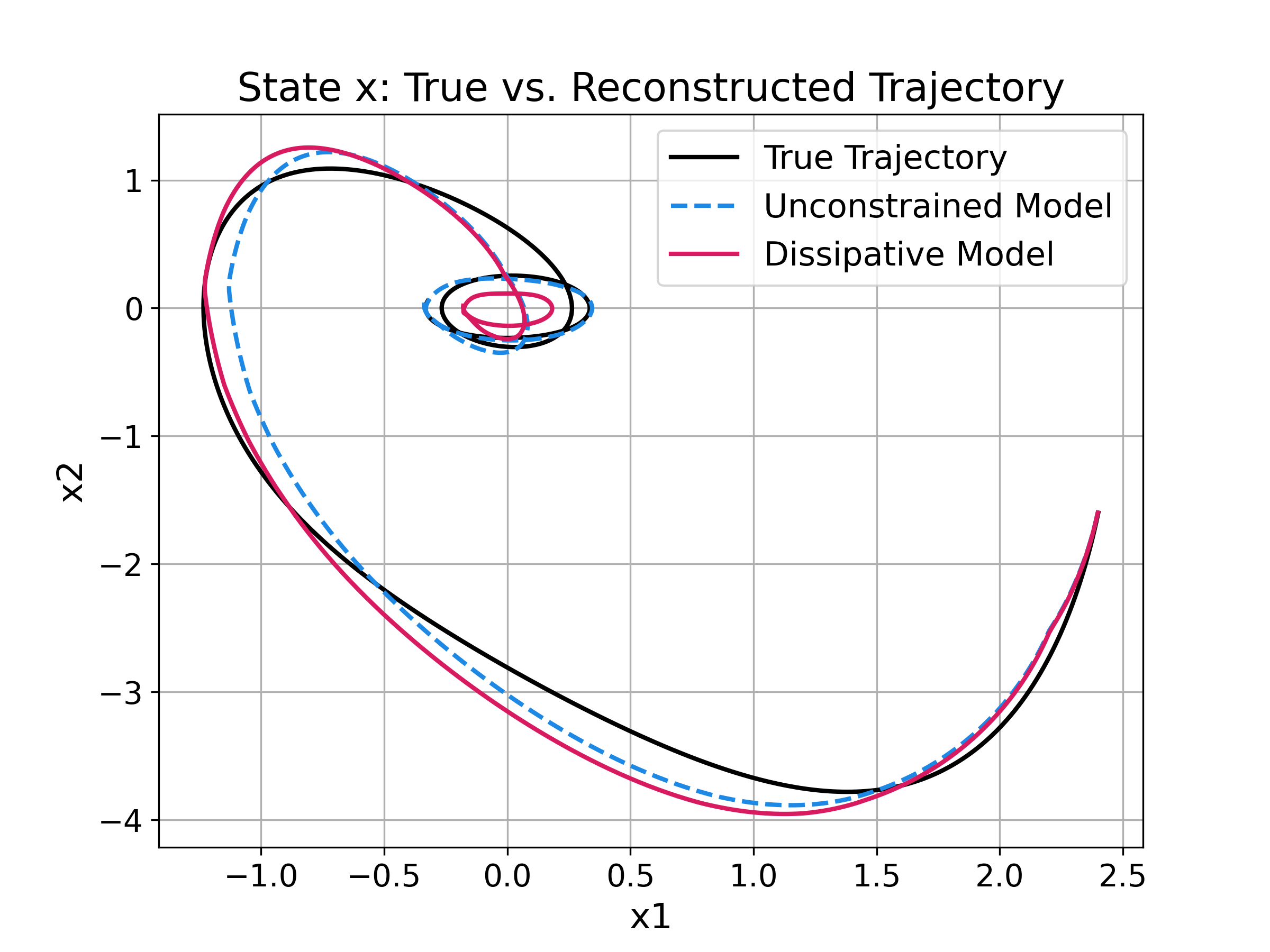}
      \caption{State trajectories of the unconstrained neural Koopman model,  the final dissipative model, and the ground truth.}
\label{fig:reconstructed_states}
\end{figure}

\begin{figure}[t] 
      \centering
      \includegraphics[scale=0.3,trim=0cm 0.8cm 0cm 0cm]{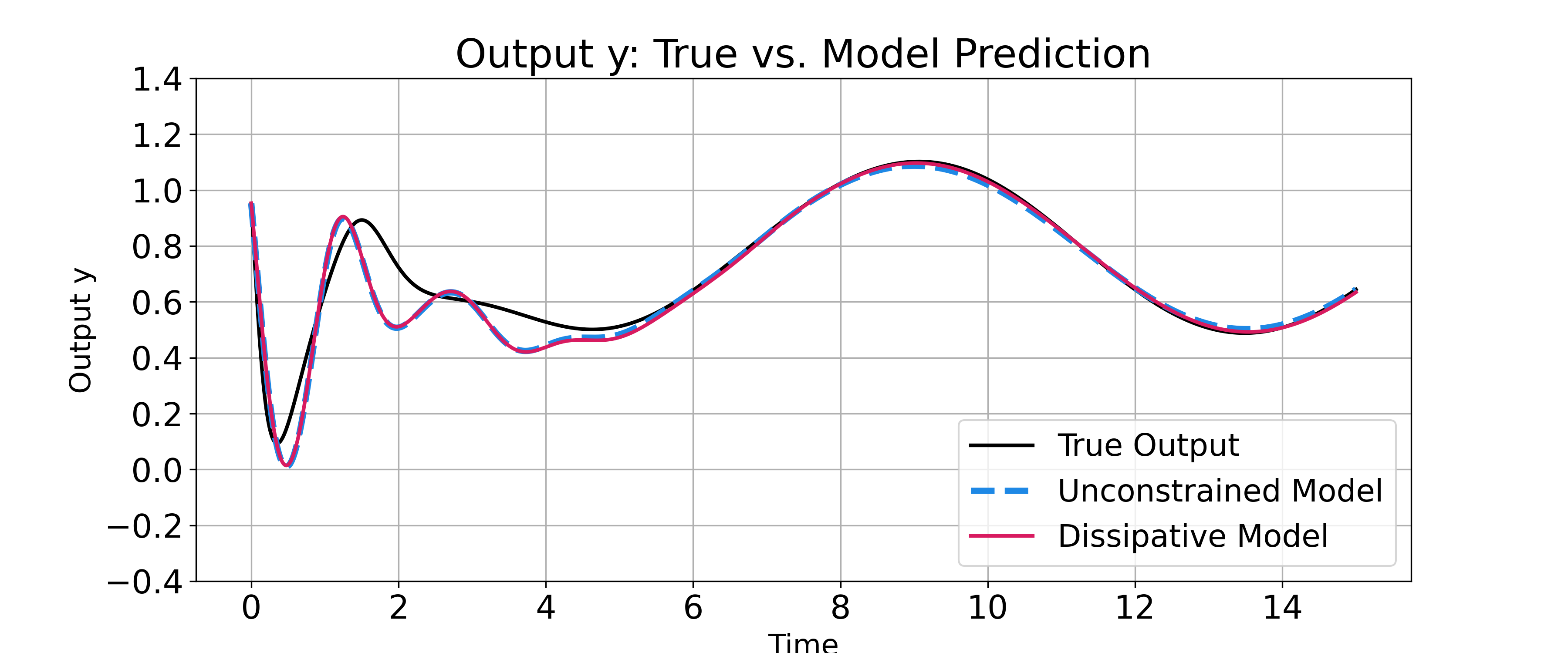}
      \caption{Model prediction of the output from unconstrained neural Koopman model,  the final dissipative model, and the ground truth.}
\label{fig:predicted_output}
\end{figure}

We illustrate the performance of our approach\footnote{The code is available at \url{https://github.com/YuezhuXu/Learning-Neural-Koopman-Operators-with-Dissipativity-Guarantee}.} on a second-order Duffing oscillator model with dynamics:
\begin{equation}\label{eqn: duffing}
    \left\{
\begin{aligned}
\dot{x}_1(t) & = x_2(t), \quad \\
\dot{x}_2(t) & = -ax_2(t)-(b+cx_1^2(t))x_1(t)+u(t),
\end{aligned}
\right.
\end{equation}
where $a,b$ and $c$ are parameters chosen as $a=1,b=1,c=1$. \\
The output function is also nonlinear, given by
\begin{equation}
y = g(x,u)= 0.2sin(0.2x_1) + 0.8cos(0.5x_2) + u
\end{equation}
This system is known to be dissipative \cite{verhoek2023convex}, and our goal is to learn a neural Koopman model  that preserves this property.  
The Koopman model is discrete as described as (\ref{eqn: linear dynamics}) and data are collected on the discretized version of system~(\ref{eqn: duffing}) as follows. 
Let $t_k = k\,\Delta t$ and $u_k := u(t_k)$ for $k=0,\ldots,N-1$. Using an explicit-Euler step with sampling period $\Delta t>0$, the discretized Duffing dynamics and measured output are
{\small\begin{equation*}
\begin{aligned}
    x_{1}(k+1) &= x_{1}(k) + \Delta t\, x_{2}(k) + w_{1}(k), \label{eq:disc-x1}\\
x_{2}(k+1) &= x_{2}(k) + \Delta t\Big(-x_{2}(k) - (1+x_{1}(k)^2)\,x_{1}(k) + u(k)\Big) + w_{2}(k),\\
y(k) &= 0.2\,\sin\!\big(0.2\,x_{1}(k)\big) + 0.8\,\cos\!\big(0.5\,x_{2}(k)\big) + u(k), 
\end{aligned}
\end{equation*}}
where $x(k) = [x_{1}(k)\;\;x_{2}(k)]^T$, and $w(k)=[w_{1}(k)\;\;w_{2}(k)]^T$ is the process noise whose $\ell_2$-norm is upper bounded by $10^{-2}$. 
In our experiments, we use $T=15$, $N=3001$, and hence $\Delta t = T/(N-1) = 0.005$.
We first learn an unconstrained model as described in Section \ref{sec: learning neural Koopman}. We excite the system using a rich set of input signals. Specifically, we use a sinusoidal excitation with an amplitude and frequency randomly selected from the range \( (0,0.5) \) and \( (0.2,0.5) \) respectively. The initial conditions are randomly generated, with each state initialized within the interval \( (-3,3) \), and data is collected from $400$ trajectories, with an $80\%/20\%$ split between training and validation trajectories. 
We choose the encoder and decoder architectures to have 2 hidden layers with 32 neurons each, and select the Koopman model dimension $N=50$. We set $\lambda_1=2,  \lambda_2=5$ and $\lambda_3=10$ for the loss function in \eqref{eqn: Koopman loss}. The initial learning rate is set to be 0.003, and the rate decays by 0.9 every 500 epochs. The maximum number of epochs is 50000. We set the initial condition to be $[2.4,-1.6]$ for testing and use test input $u(t)=0.3sin(0.7t)$. We choose a test input outside of the range of the training set to illustrate the  generalization ability of our model. As shown in Figs.~\ref{fig:reconstructed_states} and \ref{fig:predicted_output} (blue curves), the  unconstrained model accurately approximates the dynamics as both the reconstructed states and the output predicted by the model stay close to the ground truth even on unseen data. However, we note that the unconstrained model does not preserve dissipativity, despite closely approximating the system dynamics.

We now employ the perturbation approach in Section \ref{sec: alg} to enforce dissipativity. First, we use the \textit{ECLipsE} algorithm \cite{xu2025eclipse} to obtain  Lipschitz constants  $L^x_{\phi}=2.6054,L^z_{\psi}=2.6877,$ and $L^x_{\psi\circ\phi}=5.9873$. Also, by uniformly sampling 10000 points from the domain $[-4,4]^2$, we estimate the Lipschitz constant of the function $g$ to be $0.3922$. We also obtain $\phi(0)=1.4280$ and $\psi(0)=0.4190$.  Then we proceed to collect the evaluation data by sampling $500$ equally spaced values in each dimension from $[-4,4]^2$ and we obtain $\delta_c=0.1891$, $\delta_r=0.0376$, and $\delta_x=[(4-(-4))/(500-1)]/2=0.0080$ from Definition \ref{assp: data density}. Setting $\eta=1$, we perturb the system matrices by solving optimization problem \eqref{alg: perturbation} with 
with $(Q,S,R)=(-0.01,0.5,0)$,
where we pick $\lambda_A=100,\lambda_B=4,\lambda_C=1000,\lambda_D=100$, and $\tau=1$, to obtain the final $(A,B,C,D)$. Note that the perturbations from the initial matrices $A_{ini},B_{ini},C_{ini}$, and $D_{ini}$ are small - only $1.6844\times10^{-12},0.0030,1.0828\times10^{-12},2.0631\times10^{-11}$ respectively in the sense of Frobenius norm. With the same initial conditions and test input, our results  in Figs.~\ref{fig:reconstructed_states} and \ref{fig:predicted_output} (red curves) demonstrate that the final model retains fit to the system dynamical behavior,  while guaranteeing strict dissipativity in the lifted space. Importantly, from Theorem \ref{thm: dissipativity}, the dissipativity guarantee extends back to the original system. Note that the dissipative model demonstrates a slightly reduced fit near equilibrium because enforcing dissipativity essentially introduces conservatism that shrinks the reconstructed limit cycle. Away from the origin, the approximation remains close to the true system, making the tradeoff preferred for reliable closed-loop design in nonlinear regimes far from equilibrium.


\section{Conclusions}
We presented an approach to learn a neural Koopman operator model of an unknown nonlinear dynamical system that preserves the property of dissipativity. We also showed that the dissipativity of the learned model can be translated back to dissipativity guarantees on the original nonlinear system. Future work will  explore extending the framework to bilinear models for finite-dimensional approximation, adopting more general assumptions such as state-affine noise bounds, and developing more efficient sampling strategies to quantify errors with respect to unseen data.


\appendix
\textit{Proof of Theorem \ref{thm: dissipativity}.}
We choose $V(x(k))=x(k)'Px(k)$, $P=P'>0$, as the storage function in Definition \ref{def: qsr_dissipativity}. Then according to \cite{brogliato2007dissipative}, \eqref{ineq: strict dissipative} is equivalent to $\Sigma_{l}$ being strictly dissipative. We control the error between the model and the real system as follows. For any $x$, let $x_0$ be the closest point in the evaluation set \eqref{data evaluation}. The corresponding true output, output data collected, model prediction are $y_0$, $\hat{y}_0$, and $\tilde{y_0}$. With $\tilde{y}=Cz+Du=C\phi(x)+Du$, $y=g(x,u)$ and the definition of Lipschitz constant in Assumption \ref{assp: Lip on f and g}, we can write
\begin{equation*} 
{\small
    \begin{aligned}
        \|\tilde{y}-y\|\leq&\|\tilde{y}-\tilde{y_0}\|+\|\tilde{y_0}-\hat{y}_0\|+\|\hat{y_0}-y_0\|+\|y_0-y\|\\
        =& \|C\phi(x)-C\phi(x_0)\|+\|\tilde{y_0}-\hat{y}_0\|+
        \|\hat{y_0}-y_0\|+\|y_0-y\| \\
        \leq& \|C\|L^x_\phi\delta_x+\delta_r+\delta_d+L_g^x\delta_x\triangleq\delta_g(C)
    \end{aligned}}
\end{equation*}

We then bound the error on the supply function
\begin{equation*}
{\footnotesize
\begin{aligned}
        \|w(&\tilde{y},u)-w(y,u)\| =  \|(\tilde{y}-y)'Q(\tilde{y}-y)+2(\tilde{y}-y)'Q\tilde{y}+2(\tilde{y}-y)'Su\|\\
        \leq &\|(\tilde{y}-y)'Q(\tilde{y}-y)\|+2\|(\tilde{y}-y)'Q\tilde{y}\|+2\|(\tilde{y}-y)'Su \|\\
        \leq & \|(\tilde{y}-y)\|^2\|Q\|+2\|(\tilde{y}-y)\|(\|Q(C\phi(x)+Du)\|)+\tau\|\tilde{y}-y\|^2+\|S\|/\tau\\
        \leq& \delta^2_g(C)\|Q\|+2\delta_g(C)\|Q\|(\|C\|(L^x_\phi\|x\|+\phi(0))+\|D\|\|u\|)\\
        &+\tau\|\tilde{y}-y\|^2+\|S\|/\tau\\
        = & \delta^2_g(C)(\|Q\|+\tau+1)+2\delta_g(C)\|Q\|\|C\|(L^x_\phi\|x\|+\|\phi(0)\|)\\
        &+(\|QD\|^2+\|S\|^2/\tau)\|u\|^2\\
        \leq& \Bigg[\frac{\delta^2_g(C)(\|Q\|+\tau+1)+2\delta_g(C)\|Q\|\|C\|\|\phi(0)\|}{\eta^2}\\
        &+\frac{2\delta_g(C)\|Q\|\|C\|L^x_\phi}{\eta}\Bigg]\|x\|^2+(\|QD\|^2+\|S\|^2/\tau)\|u\|^2 \hspace{0pt}
    \end{aligned}}    
\end{equation*}

The last inequality holds as $\|x\|\geq \eta$. We denote the coefficient for $\|x\|^2$ in the last line as $h(C)$. Similarly, with $\tilde{x}=\psi\circ\phi(x)$, we have
\begin{equation*}
\begin{aligned}
    \|\tilde{x}-x\|\leq& \|\tilde{x}-\tilde{x_0}\|+\|\tilde{x_0}-\hat{x_0}\|+\|\hat{x_0}-x_0\|+\|x_0-x\|\\
    \leq& L^x_{\psi\circ\phi}\delta_x+ \delta_c+\delta_d+\delta_x \triangleq \delta _b
\end{aligned}
\end{equation*}
Then, we utilize $\tilde{x}=\psi(z)$ and $\|x\|\geq \eta$ to have
\begin{equation*}
    \begin{aligned}
       \|x\|\leq &\|\tilde{x}\|+\delta_b\leq L^z_\psi\|z\|+\|\psi(0)\|+\delta_b\\
       \leq& L^z_\psi\|z\|+(\|\psi(0)\|+\delta_b)\|x\|/\eta
    \end{aligned}
\end{equation*}
In other words, $\|z\|\geq (1-\frac{\|\psi(0)\|+\delta_b}{\eta}))/L^z_\psi\|x\|$.
As a result, with $\Sigma_l$ being strict dissipative with parameters $(Q,S,R,\rho,\nu)$ that satisfy \eqref{ineq: rho, nu}, we have for $\Sigma_{nl}$ that 
\begin{equation*}
\begin{aligned}
    w(y,u)\geq& w(\tilde{y},u)-(h(C)\|x\|^2+(\|QD\|^2+\|S\|^2/\tau)\|u\|^2)\\
    \geq & \rho z'z+\nu u'u-(h(C)\|x\|^2+(\|QD\|^2+\|S\|^2/\tau)\|u\|^2)\\
    \geq & \rho \left(\left(1-\frac{\|\psi(0)\|+\delta_b}{\eta}\right)/L^z_\psi\right)^2-h(C))\|x\|^2\\
    &+(\nu-\|QD\|^2+\|S\|^2/\tau)\|u\|^2 \geq 0.
\end{aligned}\\
\end{equation*}
\textit{Proof of Proposition \ref{thm: LMI}.}
Notice that \eqref{ineq: strict dissipative} is equivalent to 
\begin{equation}\label{ineq: strict dissipativity transform}
\begin{aligned}
    &\begin{bmatrix}
         P - \rho I & C'S \\
        S'C &  D' S + S' D +R - \nu I
    \end{bmatrix}\\-
    &\begin{bmatrix}
        C' \\  D'
    \end{bmatrix}(-Q)\begin{bmatrix}
        C & D
    \end{bmatrix}-
    \begin{bmatrix}
        A'\\B'
    \end{bmatrix}P\begin{bmatrix}
        A & B
    \end{bmatrix} \geq 0.
\end{aligned}
\end{equation}
If $Q<0$, we apply the Schur complement to \eqref{ineq: strict dissipativity transform} and obtain the equivalent matrix inequality
\begin{equation*}
{\footnotesize
    \begin{bmatrix}
         P - \rho I & C'S & C'\\
        S'C &  D' S + S' D +R - \nu I & D'\\
        C & D & (-Q)^{-1}
    \end{bmatrix}-
    \begin{bmatrix}
        X'\\Y'
    \end{bmatrix}P^{-1}\begin{bmatrix}
        X & Y
    \end{bmatrix} \geq 0.}
\end{equation*}
Further applying the Schur complement, we have \eqref{LMI: Q<0a}. If $Q=0$, we directly apply the Schur complement to \eqref{ineq: strict dissipativity transform} and obtain  \eqref{LMI: Q=0a}.\\
We now prove the inequalities on $\rho$. According to \eqref{ineq: rho}, the right-hand side is a monotonically increasing function of $\|C\|$. We introduce a slack variable $N_C$ such that $\|C\|\leq N_C$ and replace all the $\|C\|$ with $N_C$ in \eqref{ineq: rho}. Gathering the quadratic and linear terms in $(\delta_1+\delta_2N_C)$, we obtain
\begin{equation*}
    {\footnotesize
\begin{aligned}
    \rho \geq &\frac{(L^z_{\psi})^2(\|Q\|+\tau+1)}{(\eta-\delta_b-\|\psi(0)\|)^2} 
    \Bigg[ (\delta_1+\delta_2N_C)^2 
    \bigg(1+\frac{(2\eta L^x_\phi+\|\phi(0)\|)\|Q\|}{(\|Q\|+\tau+1)\delta_2} \bigg) \\
    &- \frac{(2\eta L^x_\phi+\|\phi(0)\|)\|Q\|}{\|Q\|+\tau+1} 
    (\delta_1+\delta_2N_C) \frac{\delta_1}{\delta_2} \Bigg]
\end{aligned}}
\end{equation*}
Note that $\delta_1=\delta_d+\delta_r+L^x_g\delta_x$, $\delta_2=L^x_\phi\delta_x$, and thus $\delta_g(C)=\delta_1+\delta_2\|C\|\leq \delta_1+\delta_2N_C$. With $c_\phi=(2\eta L^x_\phi+\|\phi(0)\|)$, we apply the Schur complement and obtain \eqref{LMI: Q<0b}, which is linear with respect to $N_C$ and $\rho$. If $Q=0$, this LMI is directly reduced to \eqref{LMI: Q=0b}. We conclude by proving the inequalities on $\nu$. The right hand side of \eqref{ineq: nu} is monotonically increasing in $\|D\|$, and we introduce slack variable $N_D$ such that $\|D\|\leq N_D$. Then, we have $\nu\geq \|Q\|^2N_D^2+\|S\|^2/\tau$. If $Q=0$, the inequality is directly reduced to  the first part of \eqref{LMI: Q=0c} (note $\|S\|/\tau$ is a constant.) If $Q<0$, applying the Schur complement, we obtain \eqref{LMI: Q<0c}. 
\\

\noindent ACKNOWLEDGMENTS: The authors thank Subhonmesh Bose (University of Illinois Urbana-Champaign) for initial discussions that led to the development of this work.

\balance

\bibliographystyle{IEEEtran}
\bibliography{ref}

\end{document}